\documentclass[manuscript]{acmart}

\AtBeginDocument{%
  }

\copyrightyear{2025}
\acmYear{2025}
\setcopyright{rightsretained}
\acmConference[CHI EA '25]{Extended Abstracts of the CHI Conference on Human Factors in Computing Systems}{April 26-May 1, 2025}{Yokohama, Japan}
\acmBooktitle{Extended Abstracts of the CHI Conference on Human Factors in Computing Systems (CHI EA '25), April 26-May 1, 2025, Yokohama, Japan}\acmDOI{10.1145/3706599.3706716}
\acmISBN{979-8-4007-1395-8/25/04}

\begin{document}

\title{Access InContext: Futuring Accessible Prototyping Tools and Methods}


\author{Patricia Piedade}
\authornote{Both authors contributed equally to this research.}
\affiliation{
  \institution{Interactive Technologies Institute, University of Lisbon}
  \city{Lisbon}
  \country{Portugal}}
\email{patricia.piedade@tecnico.ulisboa.pt}

\author{Peter A Hayton}
\authornotemark[1]
\affiliation{
    \institution{Open Lab, Newcastle University}
    \city{Newcastle}
    \country{UK}}
\email{p.a.hayton@newcastle.ac.uk}

\author{Cynthia Bennett}
\affiliation{
    \institution{Google Research}
    \city{NY}
    \country{USA}}
\email{clbennett@google.com}

\author{Anna R L Carter}
\affiliation{
    \institution{Northumbria University}
    \city{Newcastle}
    \country{UK}}
\email{anna.r.l.carter@northumbria.ac.uk}

\author{Clara Crivellaro}
\affiliation{
    \institution{Open Lab, Newcastle University}
    \city{Newcastle}
    \country{UK}}
\email{clara.crivellaro@ncl.ac.uk }

\author{Alan Dix}
\affiliation{
    \institution{Cardiff Metropolitan University}
    \city{Cardiff}
    \country{Wales}}
\email{alan@hcibook.com}

\author{Jess McGowan}
\affiliation{
    \institution{University of St Andrews}
    \city{St Andrews}
    \country{UK}}
\email{jm572@st-andrews.ac.uk}

\author{Katta Spiel}
\affiliation{
    \institution{Crip Collective || HCI Group, TU Wien}
    \city{Vienna}
    \country{Austria}}
\email{katta.spiel@tuwien.ac.at}

\author{Miriam Sturdee}
\affiliation{
    \institution{University of St Andrews}
    \city{St Andrews}
    \country{UK}}
\email{ms535@st-andrews.ac.uk}

\author{Garreth W. Tigwell}
\affiliation{
    \institution{School of Information, Rochester Institute of Technology}
    \city{NY}
    \country{USA}}
\email{garreth.w.tigwell@rit.edu}

\author{Hugo Nicolau}
\affiliation{
  \institution{Interactive Technologies Institute, University of Lisbon}
  \city{Lisbon}
  \country{Portugal}}
\email{hugo.nicolau@tecnico.ulisboa.pt}

\renewcommand{\shortauthors}{Piedade and Hayton et al.}

\begin{abstract}
The popularity of accessibility research has grown recently, improving digital inclusion for people with disabilities. However, researchers, including those who have disabilities, have attempted to include people with disabilities in all aspects of design, and they have identified a myriad of practical accessibility barriers posed by tools and methods leveraged by human-computer interaction (HCI) researchers during prototyping. To build a more inclusive technological landscape, we must question the effectiveness of existing prototyping tools and methods, repurpose/retrofit existing resources, and build new tools and methods to support the participation of both researchers and people with disabilities within the prototyping design process of novel technologies. This full-day workshop at CHI 2025 will provide a platform for HCI researchers, designers, and practitioners to discuss barriers and opportunities for creating accessible prototyping and promote hands-on ideation and fabrication exercises aimed at futuring accessible prototyping.  
\end{abstract}



\maketitle

\section{Motivation}
Accessibility has been a topic of growing interest within the CHI community, with a steady rise in publications in the last decade \cite{mack_accessibilityresearch}. Disability rights advocates within HCI, many of which are themselves disabled\footnote{We use a mix of identity-first language and person-first language throughout this article, as there are compelling arguments towards the use of both and no clear consensus within accessibility research \cite{language1,language2}.
However, personal preferences differ between individuals and should be
respected as such.} researchers \cite{hofmann_living}, have repeatedly pushed for the inclusion of disabled perspectives in work that directly or indirectly impacts them \cite{spiel_nothing_2020, mankoff_disabilitystudies}. However, research around this topic is largely unexplored, with clear gaps in HCI education \cite{elglaly_curriculum}, methodologies \cite{mack_anticipate_2022}, and tools \cite{li_accessibility_2021} that enable such participatory vision. Furthermore, though popular, accessibility continues to be treated as an afterthought within the human-centred design (HCD) process \cite{mack_anticipate_2022}. This contrasts with the known benefits of including people with disabilities in design processes \cite{shinohara_why_2021} and recent calls by disability rights advocates towards the inclusion of disabled individuals in broader settings \cite{spiel_nothing_2020, atlas_alliance_2022}. To achieve this vision of an inclusive future, and to allow people with disabilities to fully participate in design and research settings, we must ensure that these settings, practices, methods, and tools are accessible in context. This means recognising and addressing the specific barriers—whether physical, sensory, cognitive, or technological—that exist in particular settings, ensuring accessibility is not a generic afterthought but integrated into the specific contexts of design and research from the outset.

AccessSIGCHI has continuously set the precedent for creating more accessible conference experiences for disabled researchers \cite{AccessSIGCHIReport}, helping to raise the profile of research from disabled CHI authors, and showing support for this movement, while previous workshops at the ASSETS conference have begun exploring the lack of guidance in conducting accessibility research \cite{mcdonnell_tackling_2023}. Researchers with disabilities have spearheaded this movement, leveraging their unique standpoint to produce invaluable situated knowledge \cite{spiel_nothing_2020}. However, the lack of knowledge about how accessibility research should be conducted \cite{mcdonnell_tackling_2023} and the inaccessibility of existing prototyping tools and methods \cite{li_accessibility_2021, mack_anticipate_2022} means that the HCI community needs to do more to enable and promote the participation of disabled designers.

Tools and methods have been long intertwined with human ingenuity. The tools at our disposal shape our inventions—our designs. Moreover, our desire to innovate beyond what current tools may be able to shape leads to the design of new tools. The same can be said for the field of HCI, where prototyping tools are used as a foundation for design. For our workshop, we consider any digital or physical implement used in the pursuit of prototyping as a tool (e.g. this could be software like Figma, Axure RP and PowerPoint or items like pen and paper), while methods cover the context and procedures in which the tools are being used while prototyping. However, many digital tools that shape technological development are currently inaccessible to many disabled designers and researchers. For example, one study found that 55\% of GUI controls in prototyping tools had accessibility issues for screen reader users \cite{li_accessibility_2021}. Additionally, traditional low-fidelity prototyping methods such as ideation can be difficult to engage with for people with disabilities \cite{Bennett-2016, Brewer-2018}, with many visually impaired people preferring to contribute verbally to co-design activities \cite{Brewer-2018}. The HCI community is aware of these issues as previous workshops at the ASSETS conference have begun addressing the lack of guidance for how best to include people with disabilities in accessibility research \cite{mcdonnell_tackling_2023}.

Though accessibility research has begun to create a more holistic picture of accessibility challenges and of the needs of disabled people working on prototyping activities, work is still in the early stages. For example, when working on prototyping activities, people who are blind or who have low vision have preferred to use physical and tactile prototyping techniques \cite{Luebs-2024}. However, when working as a mixed ability group, people with visual impairments feel the need to split roles based on group abilities, providing more visual tasks to sighted members of a group \cite{Luebs-2024}. Additionally, neurodivergent people can and have participated in co-design activities \cite{piedade_inclusion_2024} but may struggle if, for instance, too many physical prototypes are moved around on a table surface at the same time \cite{Bennett-2016}. While digital solutions may appear to resolve issues for some, the aforementioned inaccessibility of digital tools for visually impaired people shows that more work is needed to provide novel design tools that do not simply act as a bolt-on to existing commercial tools. For instance, whilst the common high-fidelity prototyping tool Figma, as of 2023, includes some level of screen reader support for its prototype testing feature, this feature does not yet extend to the designer side of the software \cite{figma_access}, acknowledging that the accessibility gap is not being adequately addressed for disabled designers \cite{li_accessibility_2021}. Other tools, such as the digital whiteboard Miro, have made significant improvements to their WCAG compliance in order to bridge the gap between the affordances of a digital canvas for neurodivergent individuals ``to explore ideas in novel ways” and screen reader users who require adequate content descriptions and navigation features \cite{miro}. Still, some screen reader users feel self-conscious when using this form of assistive technology, particularly in workplace contexts \cite{Shinohara-2016}. It is clear that further exploration is needed to create tools that allow designers, users, and researchers with different and conflicting accessibility needs to ideate, design, and prototype together.

Building on the previous InContext efforts \cite{carter_scent_2023, carter_incontext_2022, carter_prototyping_2022}, this workshop aims to bring together a diverse group of HCI researchers, including researchers with disabilities, designers, and practitioners devoted to building more accessible futures within their discipline. Together, we will question the appropriateness of current prototyping tools and methods to propose adaptations that address accessibility challenges to bring about a more accessible technological landscape and engage in futuring novel tools and methods that embody the human-centred values of the CHI community.

\section{Workshop Structure}
In structuring the workshop, we build on the prior work of accessibility researchers \cite{spiel_nothing_2020, mcdonnell_tackling_2023, sum_dreaming, mobility_workshop, mack_anticipate_2022}; however, we also acknowledge that there is no one perfect way to run a safe, accessible, and inclusive workshop and remain open to changes based on the specific accessibility needs of participants.

Our workshop will adopt a hybrid format, taking place both synchronously in-person and online through videoconferencing and text-based platforms, as well as asynchronously so as to maximise inclusivity and access for wider audiences who may not travel to Japan and minimise the limitations of hybrid workshops, such as time zone differences.

A video-conferencing platform, such as Zoom, will be used for communication among online attendees and between online and in-person attendees. Furthermore, we will establish a conversation server, such as a Slack workspace, which will include all workshop materials to accommodate accessibility needs. All participants will be invited to access the Slack workspace before, during and after the workshop to share thoughts, resources, and artefacts. Adaptations of specific activities to hybrid format are detailed in section \ref{sec:workshopactivities}.

Given that the CHI audience is global and that some disabilities are unpredictable and require flexible timeframes for engagement \cite{mack_accessibilityresearch}, we will take particular care in ensuring that all workshop activities are facilitated and equitably accessed by those joining remotely and asynchronously. Prompts and materials for workshop activities will therefore be made available in the designated Slack workspace two days before the workshop and participants will be invited to engage in group activities and discussions. Those attending the workshop synchronously and in-person will be encouraged to interact with contributions of participants attending asynchronously to increase the likelihood of all attendees networking and engaging with one another around the workshop’s topic. Online and in-person workshop activities will be carried out in small groups of around five participants, either online or in-person, not both (see section \ref{sec:accessibility}) to ensure accessibility. Synchronous online and in-person activities will mirror each other whenever possible, with different workshop organisers serving as facilitators in the two spaces.

\subsection{Plans to Publish Workshop Artefacts}
We will publish artefacts submitted by participants in arXiv as Proceedings and on our workshop website. The arXiv publication will be made available after the workshop and include a summary of workshop discussions and outcomes. Participants will have the option to opt out of public sharing of their work. We will direct participants to CHI and SIGACCESS for guidance on writing accessible papers.

\subsection{Website}
    \label{sec:website}
The workshop web page will be hosted on the DCitizens project website and at the following link: \url{https://dcitizens.eu/access-incontext-chi25/}. All workshop information, calls for participation and participant submissions will be made available on the website, and the website maintained after the initial workshop event.

\subsection{Accessibility}
    \label{sec:accessibility}
Following the organisers' commitment to accessibility, which is inherent to the topic of this workshop, we will do our utmost to ensure the accessibility of all workshop materials and activities. All workshop content, including participant submissions, should adhere to existing accessibility guidelines set forth by CHI and SIGACCESS. All workshop group activities will be carried out in the same groups, which will only include either online or in-person participants. Though this may limit networking, we believe it will reduce accessibility barriers brought on by physically moving groups within the workshop room and synchronous hybrid collaboration. 

The tools and platforms chosen for this workshop were reviewed by the organisers to ensure the most broadly accessible choices would be made. More information regarding the accessibility of each platform will be made available on the workshop website.

Sign language interpretation and CART will be offered as a default for participants in attendance. We will gather accessibility requirements and accommodation requests from participants upon acceptance into the workshop, and procure services (e.g., additional interpreters) or make changes to the workshop activities (e.g., changing to a more accessible software) according to these. Whenever possible, we will provide participants with information regarding workshop activities ahead of time and provide reminders closer to the activity date.

\subsection{Pre-Workshop Plans (Asynchronous)}
Following acceptance notifications, we will share a workshop registration form with participants. From the form’s responses, we aim to gather information regarding accessibility needs, the number of in-person and online attendees, and consent (or lack thereof) to publically share workshop submissions. As soon as we have gathered all necessary information, we will make available on the workshop website (see section \ref{sec:website}) brief summaries of all accepted submissions and, for those who consented to public sharing of their work, full-text accessible PDFs.

Two weeks prior to the workshop, we will invite the participants to join an asynchronous conversation server (e.g. Slack). Within this setting, we will facilitate an asynchronous pre-workshop activity, encouraging participants to get to know each other in an informal setting. We will set up specific forums within the asynchronous conversation server for participants to share introductions, chat openly about the workshop or other related topics, and discuss accepted submissions.

\subsection{Workshop Activities (Synchronous)}
    \label{sec:workshopactivities}
This full-day workshop will include various activities. The specific format, activities, breaks, and lunch will be confirmed according to the CHI 2025 schedule:
\\
\emph{\textbf{Introduction \& Welcome:}}
The workshop will begin with a short introduction to the topic, participants, and facilitators. Participants will be allocated to working groups, which will remain the same during all activities, reducing any accessibility barriers, for example moving around the room and related logistics. (10 minutes)
\\
\emph{\textbf{Disability Justice Keynote:}} 
Cynthia Bennett will give a keynote talk on the disability justice framework, and the challenges of participation in collaborative prototyping faced by disabled people. This will serve to set the tone for the day and stimulate discussion about the importance of including disabled people in research (45 minutes)
\\
\emph{\textbf{Current Tools and Methodologies:}}
Participants will hold small group discussions around how they create prototypes (e.g. the tools they use) and how they consider embedding accessibility into the development of their prototypes. To encourage thinking around accessibility for different users, workshop organisers will use a set of prompts reminding groups of different access requirements (e.g. “how would a screen reader user access this”). Each group will then feedback key points from their discussion. (20 minutes)
\\
\emph{\textbf{Accessibility of tools and methodologies:}}
Building on the previous session, participants will be invited to critically question the accessibility of the prototyping tools and methods they use by considering different kinds of accessibility needs (e.g. non-visual and colourblind access, reduced distractions, dyslexia fonts). (20 minutes)
\\
\emph{\textbf{Break:}} Tea/coffee break. (30 minutes)
\\
\emph{\textbf{Brainstorming Identification:}} 
During the break, organisers will identify common topics and tools across discussions. 
\\
\emph{\textbf{Brainstorming Accessible Tools:}}
Participants will be invited to select 3 topics identified by the organisers from discussions held before the break and brainstorm how adaptations can increase the accessibility of prototyping tools and methods. Participants will then be invited to comment and build upon different groups’ ideas.
This activity will conclude with a group share-out in which participants will be asked to share their favourite idea. These ideas will be collated by the workshop organisers into an online document and this will be shared with all participants at a later date. (55 minutes)
\\
\emph{\textbf{Lunch:}} Lunch near the conference centre. (90 minutes)
\\
\emph{\textbf{Tools and Methods for the Future:}}
This will be a hands-on activity in which participants will be asked to create an accessible prototyping tool or method, which considers the topics discussed throughout the day and pushes the boundaries of what is currently technologically possible. This task is based on the potential for non-digital tools and methods being potentially more accessible and exploring this with the participants \cite{Luebs-2024}. We will provide participants with a variety of multisensory crafting materials (e.g. plasticine, pipe cleaners, paper, scissors, glue sticks, pompons, scratch-and-sniff stickers, scented markers and fuzzy felt). Workshop organisers will share a list of materials for asynchronous participants to consider working with, focused on household items. This activity will encourage participants to think beyond the computer screen to envision novel solutions that address accessibility challenges. (60 minutes)
\\
\emph{\textbf{Show, audio-describe and tell:}} Participants will be invited to show, audio-describe, and tell about their creations with the group. Participants will be encouraged to keep their objects as reminders of the need to constantly question the values embedded in the tools and methods they choose to use on a daily basis. (30 minutes)
\\ 
\emph{\textbf{Break:}} Tea/coffee break. (30 minutes)
\\
\emph{\textbf{Moving Forward:}}
Bringing together insights from the day's discussions, groups will discuss key considerations for shaping the future of accessible prototyping tools and design methods. The groups will present their ideas back to all synchronous workshop participants and a summary will be posted to Slack for asynchronous participants to read. 

Participants will then be asked to cluster these ideas to assist the development of an action plan for enabling the development of resources for accessible tools and methods for completing accessibility in context. (45 minutes)
\\
\emph{\textbf{Reflections \& Wrap Up:}}
The workshop will conclude with participants sharing any final thoughts and takeaways with each other and workshop organisers. The organisers will also take the opportunity to thank everyone for their time and share post-workshop plans (see section \ref{sec:postworkshop}). (15 minutes)
\\
\emph{\textbf{Dinner \& Further Discussions:}}
Participants will be invited to continue discussions over a workshop dinner and on the Slack workspace.

\subsection{Post-Workshop Plans (Asynchronous)}
    \label{sec:postworkshop}
Following the workshop, the Slack workspace will remain open for participants to continue discussions and build community and for organisers to share relevant updates regarding workshop outcomes. 

Organisers will work together to summarise the most relevant insights from the workshop, which will be made available to participants for discussion before any form of submission takes place. This summary may include a written component discussing the main themes and ideas emerging from the brainstorming sessions, an audio-visual library of the artefacts from the futuring exercise (including alt-text and captions for each one), and an agenda for future accessible prototyping research. Afterward, organisers will work with interested participants in creating a formal report of needs and opportunities for accessible prototyping tools and methods identified during the workshop. This write-up will be submitted for publication at an appropriate venue (e.g. ACM TACCESS, ACM ToCHI, ACM Interactions).

Aiming to reach a broader audience, we will also share our insights in a plain language blog posted on the workshop website which is part of a wider project network aiming to empower citizens with and through technology \cite{dcitizensproj22}. This blog post will be shared through the organisers’ and participants’ networks and aims to provide resources for researchers and designers in academia, industry, and non-profit organisations interested in improving the accessibility of their design processes. Our goal is to kickstart a continuous dialogue, expressing varied perspectives, extracting insights, and laying the groundwork for collaboratively creating a guiding framework that future prototyping tool designers can draw upon when navigating the complexities of this domain.

\section{Intended Audience and Recruitment}
We aim to recruit 10 - 35 participants (up to 20 in-person and 15 online participants). This number is specified to manage accessibility requirements and ensure adequate facilitation. Building upon the foundations laid by our previous InContext workshops \cite{carter_incontext_2022, carterthesis}, accessibility work \cite{Piedade23class, piedade_inclusion_2024, neto_conveying_2024, spiel19critlit, spiel_nothing_2020, Bennett18assistive, Hara15public, Roshan24aug, kokate2022exploring, dingman2021interview, Luebs-2024, li_accessibility_2021}, as well as the ongoing Horizon Europe project \cite{dcitizensproj22} we have built a network of researchers and practitioners interested in the future of prototyping, HCI tools, and technology for social good. The DCitizens project has garnered significant reach through its seminars, Discord community, mailing list, and social media channels, connecting stakeholders interested in technology for social good. Additionally, some of the co-organisers are active on social media platforms (e.g. Twitter, Mastodon, and Instagram) and members of relevant mailing lists (e.g. CHI, ASSETS). We will leverage these networks to disseminate our workshop’s call for participation, directing interested parties to our website (see section \ref{sec:website}), which will be regularly updated in the lead-up to the workshop.

\section{Call for Participation}
There has been a consistent push within HCI and Design spaces for more accessible prototyping, designs, design methods, and tools. However, practitioners struggle to find tools and methods that support these practices. Furthermore, the lack of accessible tools and methods excludes people with disabilities from fully participating in design processes, perpetuating ableism and limiting innovation.

This full-day hybrid workshop aims to explore the accessibility of prototyping tools and methods as a call to action for more inclusive practices. We strive to collaboratively investigate barriers and opportunities within current tools and jointly speculate toward a more accessible future.
 
We invite participation from researchers, designers, and practitioners across the CHI community who want to make HCI a more inclusive space. Interested parties should submit a short (1-2 page ACM camera-ready template) position paper or pictorial exploring the theme of \textit{Accessibility of Prototyping Tools and Methods: Current Realities and Future Perspectives}. Submissions should be accessible PDFs. Please contact the organisers in advance if you require assistance generating an accessible PDF. Submissions should not be anonymous.

Position papers and pictorials will be juried based on appropriateness to the call. Accepted submissions will be made available on the workshop website and in arXiv proceedings. Participants may opt out of publishing their manuscripts.
 
At least one author of each accepted submission must attend the workshop. All workshop registrants are additionally required to register for at least one day of the CHI’25 conference. 
\\
\emph{\textbf{Paper/Pictorial submissions:}} Email submissions to Patricia Piedade: patricia.piedade@tecnico.ulisboa.pt
\\
\emph{\textbf{Workshop website:}} \url{https://dcitizens.eu/access-incontext-chi25/}

\section{Organisers}
\textbf{Patricia Piedade} is a PhD student affiliated with the Interactive Technologies Institute and INESC-ID, at the University of Lisbon. Her research interests lie in accessibility and participatory methodologies, especially in the intersection of the two. Patricia's current work focuses on how to make public spaces enjoyable for neurodivergent individuals who, like herself, experience feelings of sensory overload and distress within such spaces. Patricia holds a BSc and MSc in Computer Science and Engineering both from the University of Lisbon. 

\textbf{Peter Hayton} is a visually impaired PhD student at OpenLab, Newcastle University UK. He has a strong interest in accessibility, with his research focusing on the accessibility of autonomous vehicles for visually impaired people and the accessibility of human-centred design methodologies for visually impaired researchers, such as himself. Peter has a BSc in Computer Science and has worked on digital accessibility, within the UK charity sector at a national level, before returning to academia to pursue his PhD. 

\textbf{Cynthia Bennett} is a senior research scientist at Google. Her research concerns the intersection of power, disability, design, and accessibility. She positions the lived experiences and creativity of people with disabilities as starting points for developing accessible and justice-oriented applications of AI and sociotechnical systems. She is also a disabled scholar. 

\textbf{Anna R L Carter} is a Research Fellow at Northumbria University. She has extensive experience in designing technologies for local council regeneration programs, her work focuses on creating accessible digital experiences in a variety of contexts using human-centred methods and participatory design. She works on building Digital Civics research capacities of early career researchers as part of the EU funded DCitizens Programme and on digital civics, outdoor spaces and sense of place as part of the EPSRC funded Centre for Digital Citizens. 

\textbf{Clara Crivellaro} is a Reader in Digital social justice at the School of Computing’s Open Lab, with expertise in Human-Computer Interaction, Digital Civics, Human-Centred Design, Participatory Design, and co-creation. Her research explores how the careful design of new and emergent technologies and socio-technical processes can help support democratic practices and advance equity and social justice in digital societies. She is also interested in the design of novel tools and processes to support Responsible Research and Innovation in Computing and civic-driven research commissioning processes.

\textbf{Alan J. Dix} is a Professorial Fellow at Cardiff Metropolitan University. He was elected to the ACM SIGCHI Academy in 2013 and is a Fellow of the Learned Society of Wales. Alan has worked in human–computer interaction research since the mid 1980s and is the author of one of the major international textbooks on HCI, as well as several other books including TouchIT published by OUP, which focuses on physicality and digital design, and a new volume on AI for HCI being published in 2025.  While a Luddite at heart (against technology that dehumanises), he has the hope that tools, including AI, may help designers when dealing with those with different physical and cognitive skills to themselves.

\textbf{Jess McGowan} is a PhD student in the School of Computer Science at the University of St Andrews. Their current research is focused on the potential of tabletop role playing games for better user experience design, with interests including CS education, and accessibility. Jess has a BSc in Software Development from Robert Gordon University and an MSc in Artificial Intelligence from the University of Aberdeen.

\textbf{Katta Spiel} is an Assistant Professor of Critical Access in Embodied Computing at the HCI Group of TU Wien, where they work on the intersection of Computer Science, Design and Cultural Studies. They research marginalised perspectives on technologies to inform interaction design and engineering in critical ways, so they may account for the diverse realities they operate in and in collaboration with neurodivergent and/or nonbinary peers. They are part of the Crip Collective and the EU funded ACCESSTECH project. 

\textbf{Miriam Sturdee} is a lecturer (Assistant Professor) at the University of St Andrews working on intersections of art, design, and computer science. She is a practising artist and designer and has an MFA in visual communication. Her publications explore areas of futuring, sketching and drawing, alternative research outputs, and psychology. She is particularly interested in how visual methods and outputs can be made more accessible through the application of multimodal and creative approaches.

\textbf{Garreth W. Tigwell} is an Assistant Professor in the School of Information at the Rochester Institute of Technology. His research primarily focuses on making digital content, services, and systems more accessible to disabled people by understanding and addressing challenges faced by novice and expert digital creators (e.g., mobile app and website designers). Recently, he has been exploring the role of culture in accessible design, as well as adaptable design to improve accessibility and usability for a variety of user needs and contexts of use. 

\textbf{Hugo Nicolau} is an Associate Professor at the University of Lisbon and researcher at the Interactive Technologies Institute / LARSyS. His research interests include HCI and Accessibility, focusing on the design, build, and study of computing technologies that enable positive social change. His research methods extend mostly from the discipline of HCI and are informed by perspectives in Design Justice, Psychology, Sociology, and Disability Studies. Hugo is broadly interested in research that tackles ambitious interdisciplinary problems in areas such as education, health, and social inclusion.

\begin{acks}
We thank our funding bodies for their support of this research. Patricia Piedade is funded by the Portuguese Recovery and Resilience Program (PRR), IAPMEI/ANI/FCT under Agenda no.26, C645022399-00000057 (eGamesLab), and the Foundation for Science and Technology (FCT) through scholarship 2024.00620.BD. Anna Carter and Hugo Nicolau are co-funded by the European Commission (101079116 Fostering Digital Civics Research and Innovation in Lisbon). Anna Carter, Peter Hayton and Clara Crivellaro are co-funded by EPSRC (EP/T022582/1 Centre for Digital Citizens - Next Stage Digital Economy Centre). Hugo Nicolau and Patricia Piedade are co-funded by the Foundation for Science and Technology (UIDB/50009/2020 ITI/LARSyS and UIDB/50021/2020 INESC-ID). Katta Spiel is Co-funded by the European Union (ERC, ACCESSTECH, 101117519). Views and opinions expressed are however those of the author(s) only and do not necessarily reflect those of the European Union or the European Research Council. Neither the European Union nor the granting authority can be held responsible for them. This material is based upon work supported by the National Science Foundation under Grant No. 2333220 - this NSF funding was awarded to Garreth Tigwell.
\end{acks}

\bibliographystyle{ACM-Reference-Format}
\bibliography{biblio}

\end{document}